\documentclass[journal=ancac3,manuscript=article]{achemso}

\usepackage{chemformula} 
\usepackage[T1]{fontenc} 
\usepackage{amsmath,amsfonts}
\usepackage{siunitx}
\usepackage{ulem}  

\newcommand*\Sd{\ensuremath{S_d}}
\newcommand*\Lc{\ensuremath{L_c}}
\newcommand*\sh{\ensuremath{SH_0}} 
\newcommand*\nsh{n\ensuremath{SH_0}} 
\newcommand*\pe{\ensuremath{PE}}
\newcommand*\td{\ensuremath{TD}}
\newcommand{\nm}[1]{\SI{#1}{nm}}
\newcommand{\um}[1]{\SI{#1}{\micro m}}

\author{Tristan Madeleine}
\email{tm3u18@soton.ac.uk}
\affiliation[Math]{Mathematical Sciences, University of Southampton, Southampton SO17 1BJ, United Kingdom}
\author{Nina Podoliak}
\affiliation[QLM]{Physics and Astronomy, University of Southampton, Southampton SO17 1BJ, United Kingdom}
\author{Oleksandr Buchnev}
\affiliation[ORC]{Optoelectronics Research Centre and Centre for Photonic Metamaterials, University of Southampton, Southampton SO17 1BJ, UK}
\author{Ingrid Membrillo Solis}
\author{Giampaolo D’Alessandro}
\author{Jacek Brodzki}
\affiliation[Math]{Mathematical Sciences, University of Southampton, Southampton SO17 1BJ, United Kingdom}
\author{Malgosia Kaczmarek}
\affiliation[QLM]{Physics and Astronomy, University of Southampton, Southampton SO17 1BJ, United Kingdom}

\title[Topological learning for the classification of disorder: an application to the design of metasurfaces]
  {Topological learning for the classification of disorder: an application to the design of metasurfaces}

\abbreviations{SLR,TDA}

\begin{document}


\begin{abstract}
Structural disorder can improve the optical properties of metasurfaces, whether it is emerging from some large-scale fabrication methods, or explicitly designed and built lithographically.
Correlated disorder, induced by a minimum inter-nanostructure distance or by hyperuniformity properties, is particularly beneficial in some applications such as light extraction.
We introduce numerical descriptors inspired from topology to provide quantitative measures of disorder whose universal properties make them suitable for both uncorrelated and correlated disorder, where statistical descriptors are less accurate.
We prove theoretically and experimentally the accuracy of these topological descriptors of disorder by using them to design plasmonic metasurfaces of controlled disorder, that we correlate to the strength of their surface lattice resonances.
These tools can be used for the fast and accurate design of disordered metasurfaces, or to help tuning large-scale fabrication methods. 
\end{abstract}


Metasurfaces are two-dimensional metamaterials with subwavelength scattering elements designed to have electromagnetic properties unobtainable from bulk materials\cite{chen_review_2016}. 
However, their designs usually require expensive and time consuming fabrication methods, such as lithography, limiting their large-scale and large surface area production.
In order to circumvent these limitations, extensive work is being conducted to devise quicker and less expensive fabrication methods for large-scale metasurfaces, usually at the cost of emerging structural disorder. Some examples are gas-phase cluster beam deposition\cite{mao_single-step-fabricated_2021}, nanosphere photolithography\cite{ushkov_nanosphere_2021} or lithography free fabrication methods\cite{yildirim_disordered_2019}, such as bottom up self assembled systems\cite{wu_disordered_2022,tani_transparent_2014,chen_visible_2019,piechulla_antireflective_2021,shutsko_light-controlled_2022}, colloid deposition\cite{piechulla_fabrication_2018} or  polymer phase separation\cite{narasimhan_bioinspired_2020,siddique_bioinspired_2017,donie_phase-separated_2021}.

While the emergence of structural disorder is usually thought as being an unavoidable downside of these fabrication methods, some photonic based applications actually benefit from it \cite{cao_harnessing_2022}.  Indeed positional disorder helps to tune \cite{el_shamy_light_2019,antosiewicz_localized_2015,sterl_shaping_2021} or reduce the diffraction\cite{chevalier_absorbing_2015}, scattering\cite{zhang_hyperuniform_2021}, reflection \cite{tani_transparent_2014,piechulla_antireflective_2021} or radiation \cite{sterl_shaping_2021,haghtalab_ultrahigh_2020} of metasurfaces, with potential applications in the fabrication of better displays \cite{bertin_correlated_2018}.  Disorder can also suppress grating effects\cite{sterl_design_2020} , make surface-enhanced Raman scattering broadband\cite{narasimhan_bioinspired_2020}, enhance localised photoluminescence \cite{roubaud_far-field_2020},improve wavefront shaping \cite{veksler_multiple_2015,jang_wavefront_2018}, improve light absorption\cite{kim_hyperuniform_2021,reyes-coronado_enhancement_2022},e.g. for solar cells\cite{siddique_bioinspired_2017}, or light extraction \cite{jouanin_designer_2016,wu_disordered_2022}.
For example, coating the air-LED interface with disordered nanostructures provides a broadband coupling between what would have been internally trapped photons to the external radiation, making more energy efficient LEDs\cite{mao_single-step-fabricated_2021}. 
In some of these fields, correlated disorder seems to be particularly important.
Indeed, a correlation length, either induced by a minimum distance between the nanostructures, or by some stealthy hyperuniformity properties helps to create metasurfaces with broader absorption bands\cite{kim_hyperuniform_2021}, broader diffusive properties \cite{zhang_hyperuniform_2021} or prevent light trapping between nanostructures for more efficient light extraction\cite{jouanin_designer_2016}. 

The different applications of disordered metasurfaces lead to more recent effort to tailor disorder for specific desired optical properties \cite{bertolotti_designing_2018,rothammer_tailored_2021,dupre_design_2018,yu_engineered_2021}, for example using  inverse design methods \cite{pestourie_efficient_2023,li_empowering_2022} based on machine learning \cite{jiang_deep_2021,khoram_graph_2023} or via topology optimisation \cite{hammond_phase-injected_2023,ballew_constraining_2023}.
Indeed, combining disorder engineering and topology optimisation one can build  metasurfaces with selective light polarisation conversion while minimising the in-plane phase fluctuation \cite{xu_emerging_2022}.
While these methods directly generate optimised disordered patterns, they can be time consuming and computationally expensive to implement.
In some cases, knowing the link between disorder and the optical properties of a metasurface could significantly speed up the design process by restricting the optimisation to the degree of disorder of a metasurface.
However, despite many methods existing to quantify disorder, they all have their strength and weaknesses and are only relevant for specific applications \cite{yu_engineered_2021}.
In this work, we present new topology inspired numerical tools suitable for the characterisation of disordered metasurfaces. 
Their universality makes them useful for the characterisation of both correlated and uncorrelated disorder, and can be used either for the characterisation of disordered metasurfaces built with techniques similar to those mentioned above, or for the fast and accurate design of metasurfaces of specific disorder levels.
We demonstrate the relevance of these tools by designing and fabricating  metasurfaces made of plasmonic nanostructures embedded in dielectric media whose structural correlated disorder is related to the strength of their surface lattice resonances (SLR).

\section{Results and discussion}
We first present a generalised model of disorder generation.  We show that a large correlation length may lead to potentially ambiguous designs, where the degree of disorder is poorly represented by the generative/statistical parameters, hinting to the need for better disorder descriptors. We then introduce the field of Topological Data Analysis (TDA) and the tools required to characterise metasurfaces.  Using them, we show that correlated disordered metasurfaces are poorly represented by their generative parameters, while being suitably described by these topological descriptors.  We then prove the characterisation accuracy and predictive properties of these tools by designing metasurfaces with specific disorder levels, first theoretically then experimentally.

\subsection{Models of correlated and uncorrelated disorder}

\begin{figure}[ht]
  \begin{center}
    \includegraphics[width=0.65\linewidth]{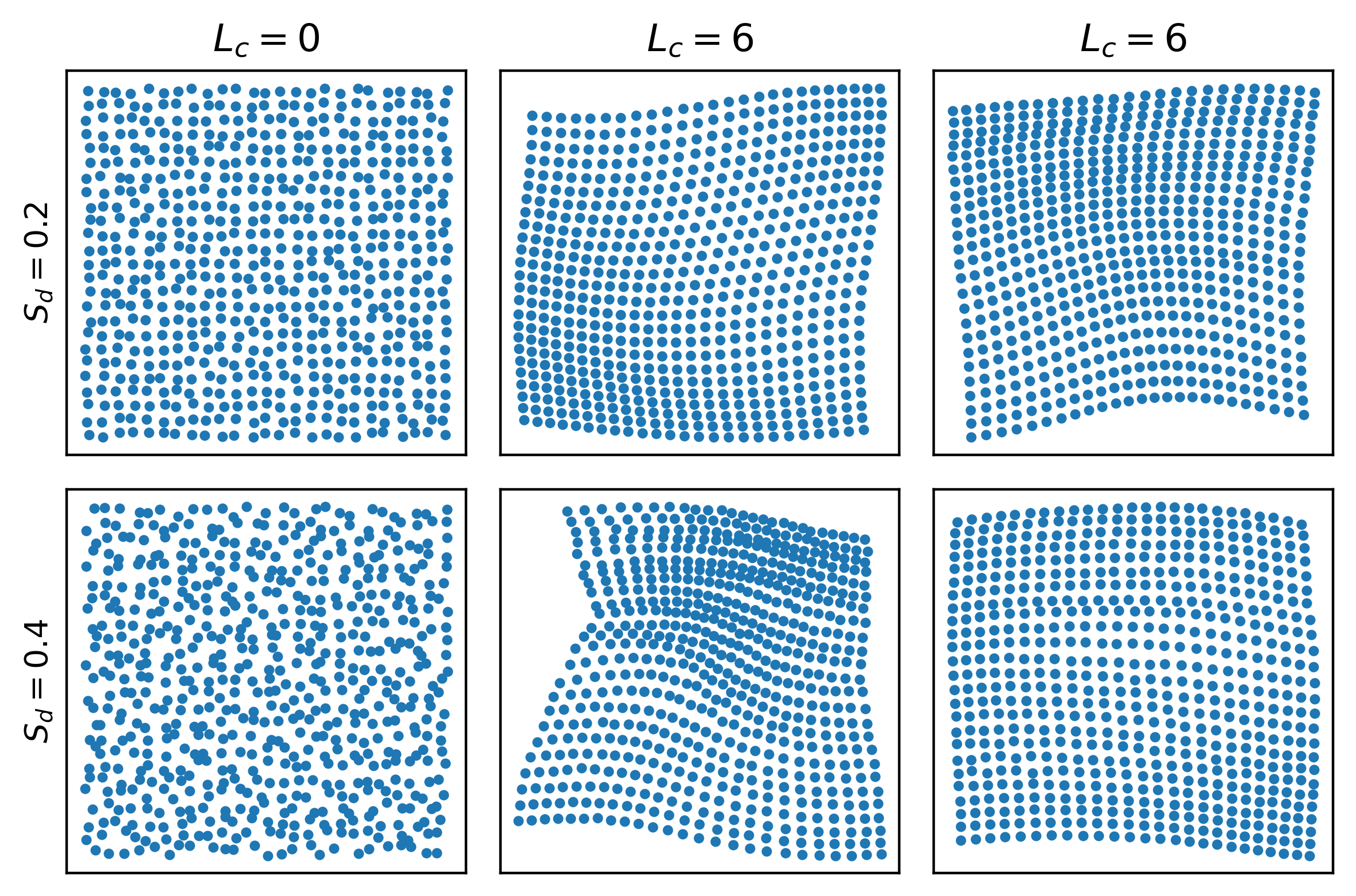} 
    \caption{ Examples of generated disordered lattices.  The top row corresponds to lattices generated with \Sd{}=0.2 and the bottom row correspond to lattices generated with \Sd{}=0.4.  The left column represents uncorrelated disordered lattices, \Lc=0, while the middle and right column represents correlated disordered lattices, \Lc{}=6.}
    \label{fig:Visual_ambiguity}
  \end{center}
\end{figure}

A recent work\cite{sterl_shaping_2021} presented a model of disorder to study how correlated and uncorrelated disorder influences the far-field optical response of a metasurface.
We apply our characterisation of disorder on their model that we reintroduce here.

Starting with a regular lattice, such as a square lattice of period $P$ made of $N_x \times N_y =N$ nanostructures whose positions are defined by $\overrightarrow{r_i}, i \in \left[1,N \right]$, we can define correlated and uncorrelated disorder as follows.  Each nanostructure position is modified by a random vector $\overrightarrow{\Delta r_i}$ whose $x$ and $y$ components are generated from a continuous uniform probability distribution bounded by $\left[-S_d P, S_d P \right]$.  The non dimensional parameter \Sd{} determines the strength of the uncorrelated disorder.  A correlation length can be implemented by adding to $\overrightarrow{\Delta r_i}$ the uncorrelated disorder of nearby nanostructures, indexed by $j$, weighted according to how far they are with a factor $C_{ij}$,
\begin{equation}
  C_{ij}= e^{-\left(\frac{r_{ij}}{2 L_c P} \right)^2},
\end{equation}
with $r_{ij}$ the distance between the nanostructure $i$ and $j$, including the uncorrelated disorder applied to $\overrightarrow{r_i}$ and $\overrightarrow{r_j}$.  The correlation length is given by the full width at half maximum of $C_{ij}$ which is equal to $2 \sqrt{2 \ln 2} L_c P$, which is proportional to the non dimensional parameter \Lc{}.  The total disorder perturbation can be summarised as:
\begin{equation}
  \label{eq:disorder_formula}
  \overrightarrow{r_i}' = \overrightarrow{r_i}
  + \overrightarrow{\Delta r_i}
  + \sum \limits_{j \neq i} \overrightarrow{\Delta r_j} C_{ij}.
\end{equation}
Using the expression~\eqref{eq:disorder_formula}, we generated lattices with varying values of \Lc{} and \Sd, see figure~\ref{fig:Visual_ambiguity}.  On the lef column of figure~\ref{fig:Visual_ambiguity}, one can visually appreciate that the strength of uncorrelated disorder, for \Lc$=0$, is well represented by \Sd.  Adding a correlation length, for example in the middle and right columns of figure~\ref{fig:Visual_ambiguity} makes the lattices' distortion smoother.  However, if one can still guess that the middle column, bottom lattice is more disordered than the middle column, top lattice, as they were respectively generated with \Sd$=0.4$ and \Sd$=0.2$, this is not systematically the case.  Indeed, the lattices in the right column were generated with the same parameters \Lc{} and \Sd{} as in the middle column, however the relative disorder strength of these two lattices is ambiguous.

While a correlation length makes the generative parameter \Sd{} less accurate to represent the positional disorder of a lattice, it also destroys the information about the original regular lattice by inducing collective movements of the lattice's points.  This makes a statistical description of correlated disordered lattices much harder due to not having a reference lattice to compare them to.  In order to circumvent these constraints, we introduce topology inspired numerical tools allowing to compare lattices with each other in a way that highlights the influence of a correlation length and provides a more accurate measure of disorder than \Sd{}.

\subsection{Topological characterisation of disorder}
\label{ssec:tda_disorder}

\begin{figure}[ht]
  \begin{center}
    \includegraphics[width=0.95\linewidth]{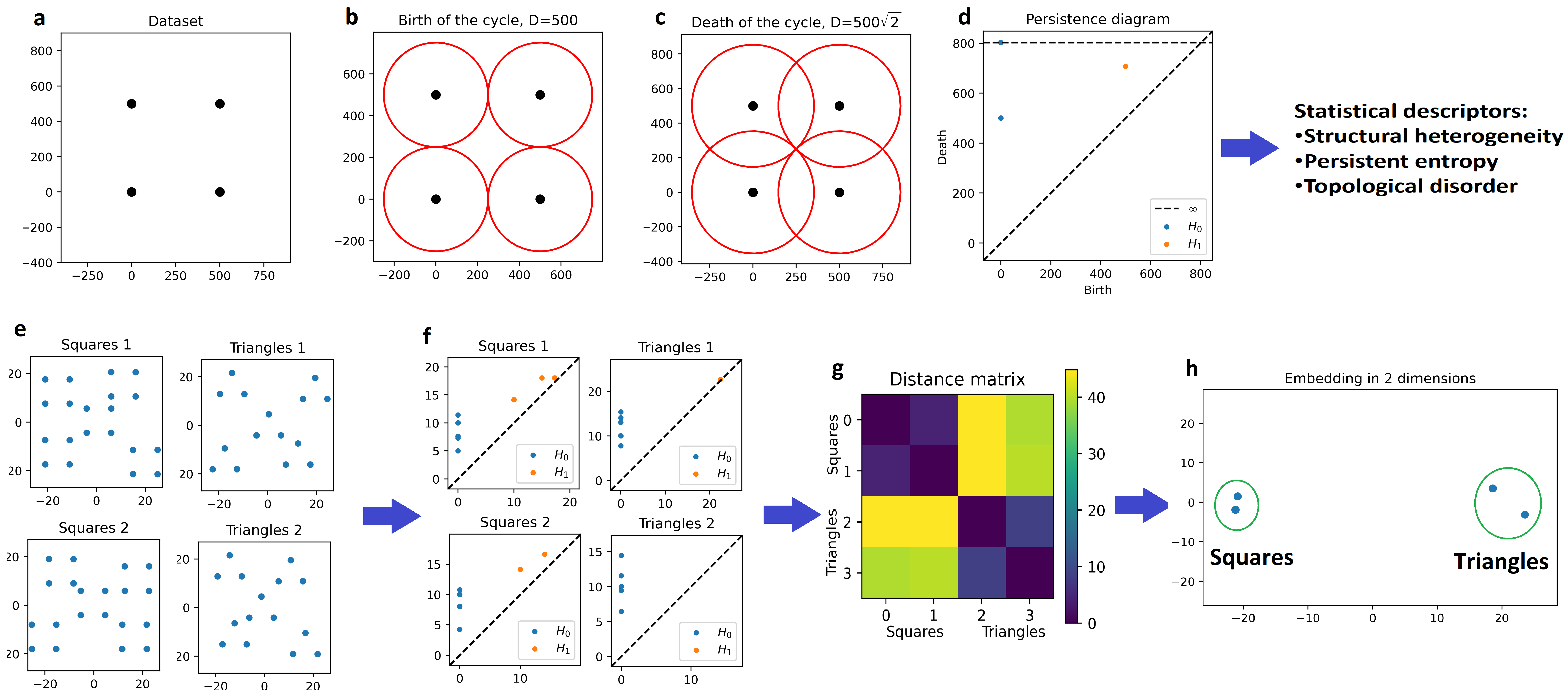} 
    \caption{Examples of two key TDA process used in this paper. The top row represents the computation of persistent homology from the dataset (a) to its representation in a persistent diagram (d). In (b) and (c) are represented the circles whose diameters correspond respectively to the birth and death of the loop of this dataset (single $H_{1}$ point in the persistence diagram in panel~d).  The bottom row represents the computation of the embedding of datasets (e) in a two dimensional space (h) via the computation of their persistence diagrams (f) and the distance between them (g).  Datasets of the same type are clustered in the embedding space (panel~h).
    \label{fig:TDA} }
  \end{center}
\end{figure}

Topological Data Analysis (TDA) is a collection of tools inspired from topology and geometry designed to provide qualitative and quantitative descriptors of structures in datasets.  It has had various applications in different fields ranging from cosmology\cite{xu_finding_2019,heydenreich_persistent_2021} to solid state physics \cite{cramer_pedersen_evolution_2020,hiraoka_hierarchical_2016,ormrod_morley_persistent_2021,hirata_structural_2020,onodera_structure_2020}.  One of its sub-fields, persistent homology, is particularly efficient to recover the scale of topological features.  We provide here a brief description of the procedure. Interested readers can find detailed introductory notes\cite{edelsbrunner_computational_2022,chazal_introduction_2021} and the whole process can be executed using standard libraries such as GUDHI\cite{gudhi:urm} or Ripser\cite{Bauer2021Ripser}.

Starting from a point cloud such as that in panel a of figure~\ref{fig:Visual_ambiguity}, we build a collection of topological spaces called Rips simplicial complexes, indexed by a real number $r$. For a given value of $r$, the complex is constructed as follows. A ball of radius $r$ is drawn around each point of the point cloud. If two balls intersect, we add a link between their respective centres. Similarly, higher order links are added to the complex upon the intersection of three or more balls.  Restricting ourselves to only two dimensions, which is relevant for flat metasurfaces, we draw circles of radius $r$ around each point and we only consider connections between pairs and triplets of points.  The topological properties of each simplicial complex, the number of connected components and the number of loops in two dimensions, can be directly computed using algebraic topology.  Tracking the evolution of these topological features for different values of $r$ provide useful insight over their scale.  These features can be summarise in a persistence diagram for which each feature, indexed by the integer $i$, is represented by two coordinates, their "birth", $b_i$ and their "death", $d_i$, which are the values of $2r$ at which they appear and disappear.

For example, if we consider a simple point cloud such as a square of side 500 in figure~\ref{fig:TDA}a, we see the birth of a loop when the diameter of the circles is equal to the side length of the square, figure~\ref{fig:TDA}b. This loop dies when the diameter is equal to the diagonal of the square, figure~\ref{fig:TDA}c, and the birth and death of this loop is represented as the point at coordinates $(500,707)$ in a persistence diagram, figure~\ref{fig:TDA}d. Additionally, four connected components are born at $r=0$ and three of them die such that only one remains after the circles intersect in figure~\ref{fig:TDA}b. Therefore, three (overlapping) points at coordinates $(0,500)$ are represented in figure~\ref{fig:TDA}d. The last connected components remains for $r\rightarrow \infty$. As we stopped the computation of persistent homology at $r=400$ we assign to this point coordinates $(0,800)$.

\begin{figure}
  \centering
  \includegraphics[width=0.9\linewidth]{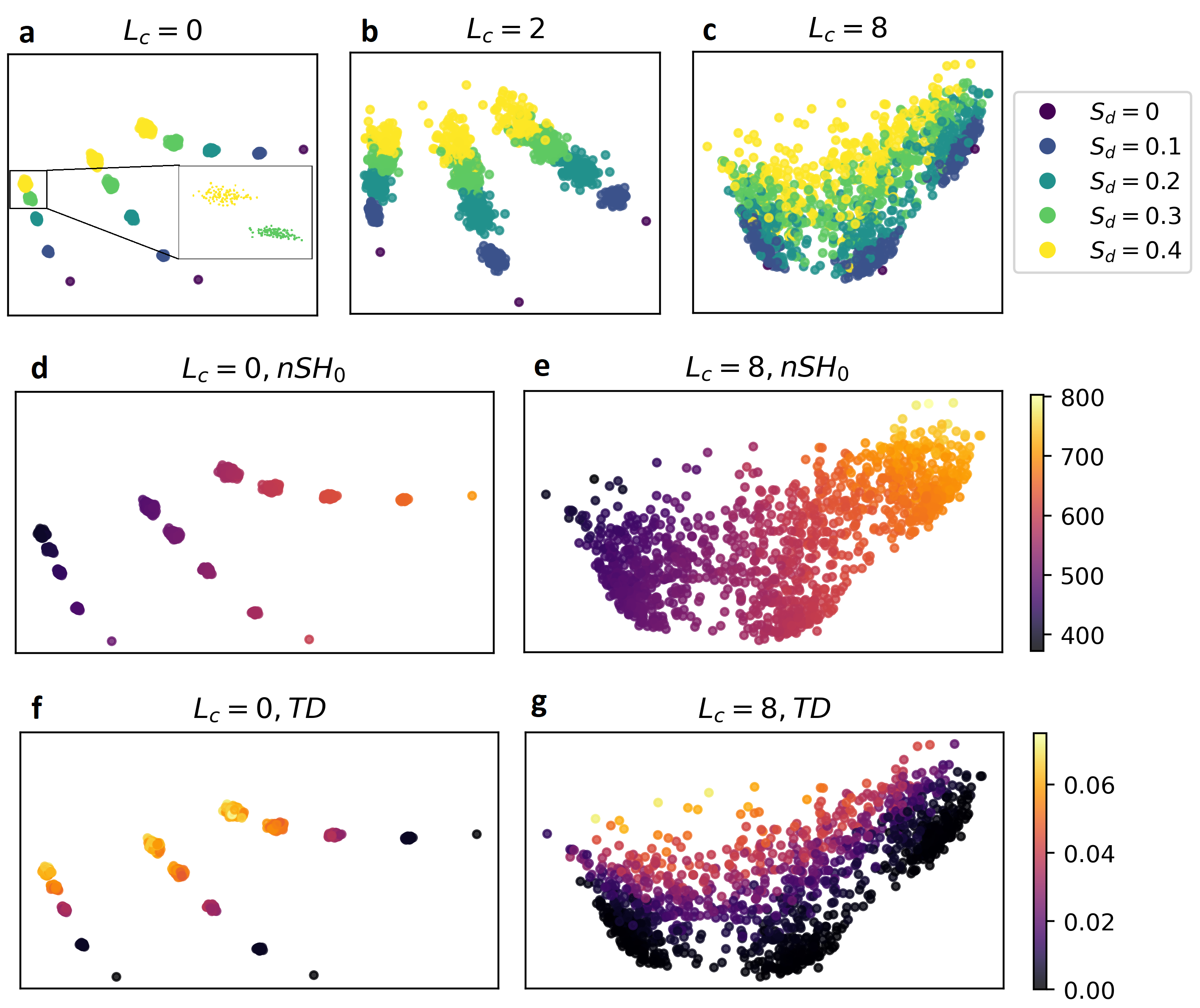} 
  \caption{
    \label{fig:Embedding}
      Scatter plots of the two dimensional embedding of three sets of generated lattices with uncorrelated (a), weekly correlated (b) and strongly correlated (c) disorder. Each set was generated from an original square lattice of period 500, 600 and \nm{700} (left to right in panel a) and with $\Sd{} \in \left[0,0.4\right]$.  In the absence of correlation, lattices with different values of the period and of \Sd{} are well clustered. In the insert of panel~a we adapted the size of points to illustrate how clustered the lattices are. The clustering is lost in the presence of correlations, panels~b and~c. Panels~d and~f, and panels e and g are equivalent to panels a and c respectively, with colour coding based on the value of n\sh~(\td) in panels d and f (e and g). In both cases the colour gradient is not significantly affected by correlation.}
\end{figure}

The second row of figure~\ref{fig:Embedding} illustrates how TDA can be used for the clustering analysis of point clouds. The first step is to measure the ``distance'' between point clouds (figure~\ref{fig:Embedding}e).
We do this by measuring the distance between their corresponding persistence diagrams.
Several metrics can be defined over the space of the persistence diagrams and we have selected to use the Wasserstein distance\cite{edelsbrunner_topological_2000} for its simplicity of use (figure~\ref{fig:Embedding}f and~\ref{fig:Embedding}g).
If several point clouds are considered, one can build a geometrical embedding, for example via classical multidimensional scaling\cite{wang_classical_2012}, in which each point cloud can be represented as one point and the distance between each point is given by the distance between their respective persistence diagrams (figure~\ref{fig:Embedding}h).
This provides a visual representation of the configuration space of the different point clouds and can be used to detect clustering.  For example, we considered two sets of four point clouds made of either triangles or squares, such as represented in figure~\ref{fig:TDA}e. 
Upon computing their persistence diagrams,figure~\ref{fig:TDA}f, and the distance matrix between then, figure~\ref{fig:TDA}g, we observe that a square seems to be more similar, or closer, to the other square than to the triangles. 
This can be directly visualised in their embedding in figure~\ref{fig:TDA}h, where we observe two clusters corresponding to the sets of squares and triangles.

In order to visualise the space of configurations obtained from the definition of correlated disordered lattices in \eqref{eq:disorder_formula}, we performed the embedding of 1203 lattices generated for three different lattice periods, 500, 600 and \nm{700}, and five different values of $\Sd{} \in \left[0,0.4\right]$.  We repeated this for uncorrelated disorder, \Lc{}=0, weakly correlated disorder, \Lc{}=2, and strongly correlated disorder, \Lc{}=8 (panels a,b and c of figure~\ref{fig:Embedding}).  
We see in figure~\ref{fig:Embedding}a an unambiguous clustering of the uncorrelated disorder lattices generated with the same parameters, i.e. \Sd{} and the original lattice period.  For a fixed value of the period, the lattices appear to live on a simple curve on which we see five separated clusters of points corresponding to the five values of \Sd{} considered.  As can be seen in figure~\ref{fig:Embedding}b, a correlation length increases the size of each cluster, allowing them to overlap.  For a large correlation length, like in figure~\ref{fig:Embedding}c, each cluster is so large that they form one common cluster between the three different periodicities and the different values of \Sd{}.  This is in agreement with what we presented in the previous section. 
A non-zero correlation length allows the lattices to explore a broader configuration space.  Therefore, one can generate lattices that are more different from each other despite using the same parameters and the overall size of the clusters, related to the maximum distance between the persistence diagrams of the generated lattices, is bigger.  This expansion of the lattices configuration space leads to situations where two lattices generated with different parameters may actually be very similar to each other, which is represented by the overlap between the clusters in figure~\ref{fig:Embedding}b.  Eventually, a large enough correlation length will make the clusters big enough to make the generative parameters obsolete, which is what we observe in figure~\ref{fig:Embedding}c.  Indeed, even if one can see some general trend between the overall lattices position and the value of \Sd{}, one cannot accurately recover the value of \Sd{} based on the lattice position.  This can lead to situations where a lattice generated with a high amount of disorder, i.e. a large value of \Sd{}, may be as, or more, ordered than a lattice generated with a small amount of disorder, such as represented in the right column of figure~\ref{fig:Visual_ambiguity}.

Using TDA, we were able to show the limitation of the parameter \Sd{} to characterise generated lattices.
One can build several metrics to describe persistence diagrams, which can be used as simpler descriptors of the topology of datasets, or as inputs of more refined machine learning based models \cite{hensel_survey_2021,leykam_topological_2023}.
In this work, we use two statistical descriptors based on lattices' persistence diagrams in order to describe both the typical distance between each point of the lattices and their positional disorder.
The first numerical descriptor is normalised structural heterogeneity of degree 0 (n\sh{}) and is the sum of the lifetime of the topological features of degree 0, the connected components\cite{membrillo_solis_tracking_2022}, divided by the number of points of the lattice, $N$: 
\begin{equation}
  \label{eq:1}
  n\sh{} = \frac{1}{N}\sum \limits_{(b,d) \in H_0 \subset \mathfrak{D}}  d-b, 
\end{equation}
with $b$ and $d$ the birth and death of each topological feature of degree 0, $H_0$, of the persistence diagram $\mathfrak{D}$.
As the death of the topological features of degree 0 is proportional to the distance between the points of the lattices, n\sh{} can be directly related to the average nearest neighbour distance between the nanostructures.
If we colour the embeddings of uncorrelated and strongly correlated, \Lc{}=8, lattices of figure~\ref{fig:TDA} according to the value of \nsh{} of each lattice, we see in figure~\ref{fig:TDA}d that this quantity almost recovers perfectly the periodicity of the lattice for uncorrelated disorder, which confirms our interpretation of the topological features of degree 0.  When applied to strongly correlated disordered lattices, figure~\ref{fig:TDA}e, \sh{} provides a smooth ordering of the lattices, following a similar trend as for uncorrelated disordered lattices.

We also introduce a new descriptor that we call Topological Disorder (\td{}), inspired from the persistent entropy (\pe{}) \cite{atienza_new_2018, he_persistent_2022, ali_survey_2022}. \pe{} is defined as
\begin{equation}
  \label{eq:2}
  \pe = - \sum \limits_{(b,d) \in \mathfrak{D}} \frac{d-b}{L} \ln \left( \frac{d-b}{L} \right),
  \qquad
  L = \sum  \limits_{(b,d) \in \mathfrak{D}} d-b.
\end{equation}
\pe{} is maximal for $d-b=l \text{ (constant)}, \forall (b,d) \in \mathfrak{D}$ and equal to $\log{\Omega}$, with $\Omega$ the total number of topological features in $\mathfrak{D}$.  Therefore, \pe{} is maximal for regular, periodic lattices and measures how ordered lattices are.  In order to avoid the counter intuitive association of a highly ordered lattice with its high persistent entropy, and to define a measure of disorder independent of the lattice's size, which modifies the number of topological features $\Omega$, we define \td{} as
\begin{equation}
  \label{eq:TD}
  \td{}= \sum \limits_{i} \left[  1 + \frac{\sum \limits_{(b,d) \in H_i \subset \mathfrak{D}} \frac{d-b}{L_i} \ln \left( \frac{d-b}{L_i} \right)}{\log \Omega_i} \right]  ,
  \qquad
  L_i = \sum  \limits_{(b,d) \in H_i \subset \mathfrak{D}} d-b,
\end{equation}
where we split the computation over the degrees $i$ of the topological features, in order to capture the fundamental differences between topological features of different degree.
Indeed, one can see on figure~\ref{fig:TDA}d that, despite the regularity of the dataset in figure~\ref{fig:TDA}a, the topological features in the persistence diagram are located in different places, which would artificially increase the value of \td{}.
While the example in figure~\ref{fig:TDA}a is simple, this remains the case for ordered lattices.
By construction, \td{} is invariant by rescaling of the typical length of the lattices, making it an orthogonal descriptor of the lattices with respect to \nsh{}.  \td{} is also minimal for ordered lattices, equal to 0, and is independent of the number of points of the lattices.  Therefore, it can be used as a universal measure of disorder, not only for point clouds perturbed from different periodic lattices array, but also for point clouds without any inherent order, such as in self-assembled systems. 
If we colour the embeddings of uncorrelated and strongly correlated lattices of figure~\ref{fig:Embedding} according to their \td{}, we see in figure~\ref{fig:Embedding}f that \td{} recovers perfectly the strength of the uncorrelated disorder, regardless of the lattices' periodicity, which confirms that \td{} is indeed a measure of the lattices' disorder.  When applied to strongly correlated disordered lattices, figure~\ref{fig:Embedding}g, \td{} provides another smooth ordering of the lattices, orthogonal to the one given by \nsh{}.

These observations suggest  that \td{} and \nsh{} are two topologically inspired descriptors that can be used to quantify the positional disorder and the typical distance between points of a dataset respectively.  Being, by construction, independent of any reference dataset, these tools are suitable to classify datasets that are not easily described using classical statistical methods, such as correlated disorder point clouds or self-assembled systems.

\subsection{Tailored metasurface design, fabrication and spectroscopy} \label{ssec:theory}

\begin{figure}[htbp]
  \begin{center}
    \includegraphics[width=0.9\linewidth]{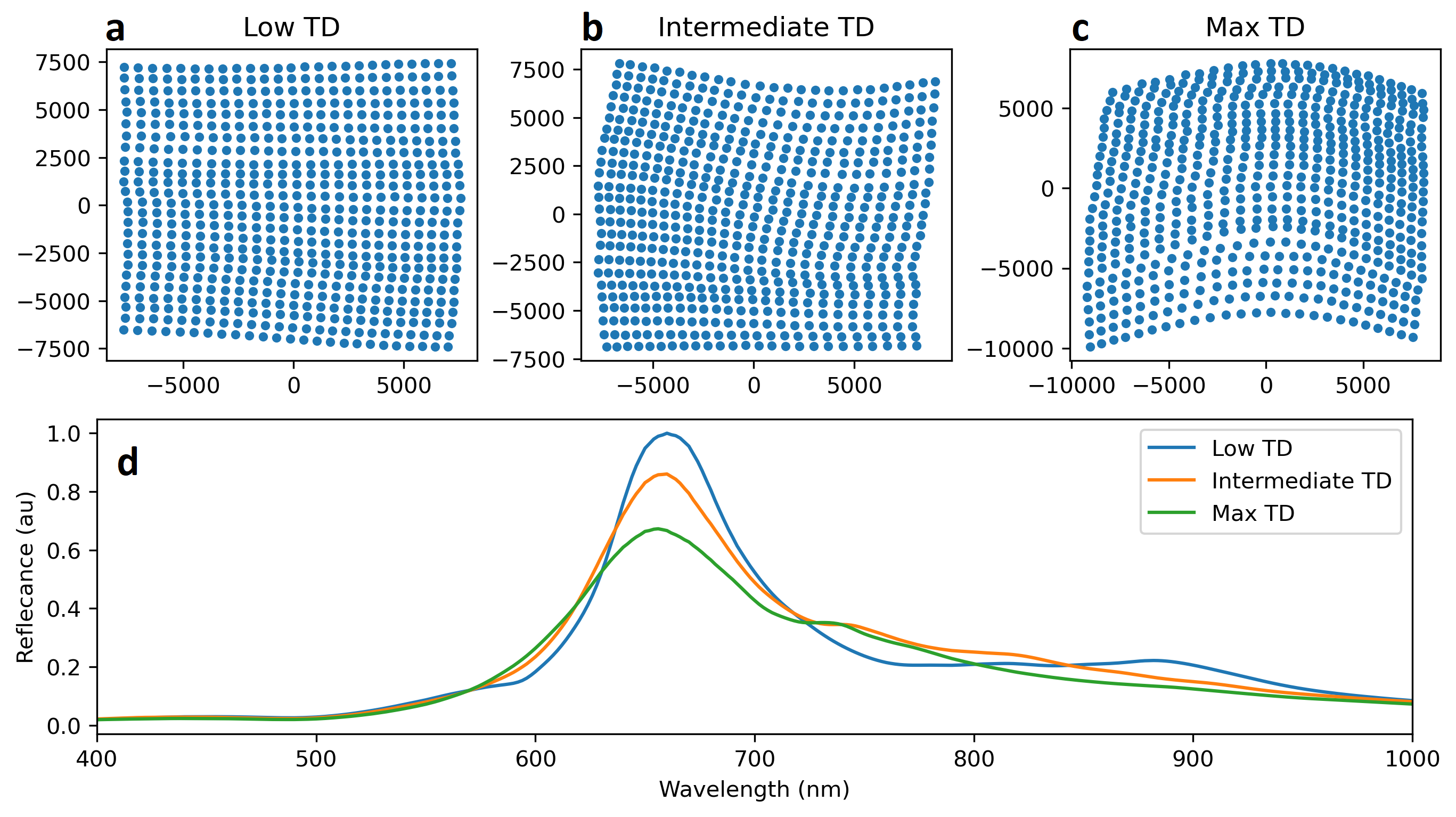} 
    \caption{Theoretical investigation of the correlation between \td{} and the strength of SLR. a,b and c represent respectively the generated metasurfaces of lowest, median and highest \td{}. Their computed reflectance spectrum, in arbitrary units, is represented in d.} \label{fig:model}
  \end{center}
\end{figure}

We demonstrate the accuracy of \td{} by using it to design, and subsequently build, plasmonic metasurfaces of specific degree of disorder, that we relate to the strength of their SLRs.  We first investigate the link between \td{} and the strength of the SLRs theoretically using the discrete dipole approximation\cite{fradkin_fourier_2019}. 
We randomly generated lattices of 25$\times$25 points with \Lc{}=8 and \Sd{}=0.3, starting from a square lattice of period \nm{500}, where each point represents the position of a plasmonic nanostructure.  Filtering the point clouds using \nsh{}, we restrict ourselves to metasurfaces of similar nearest neighbour distance. 
From these point clouds, we pick those with the highest, lowest and median value of \td{} (figure~\ref{fig:model}a, b and c).  We consider each nanostructure to be a gold nanocylinder of height \nm{50} and diameter \nm{120} whose optical properties, under the dipole approximation, are fully determined by their polarisability.  The gold nanocylinders are assumed to be embedded in an homogeneous glass like dielectric layer of refractive index 1.41.  We numerically compute the reflectance of the three metasurfaces under illumination by a circularly polarised plane at normal incidence, figure~\ref{fig:model}d. 
As predicted, the higher the  topological disorder, the weaker the SLRs are. Indeed, we can see on figure~\ref{fig:model}d that the amplitude of SLR dip is inversely proportional to \td{}.  Similarly,  the quality factors of these resonances which are 8.2, 7.5 and 6.5 for the lowest, median and highest \td{} respectively.

\subsection{Experimental verification of the \td{}--SLR link }

\begin{figure}
  \begin{center}
    \includegraphics[width=0.95\linewidth]{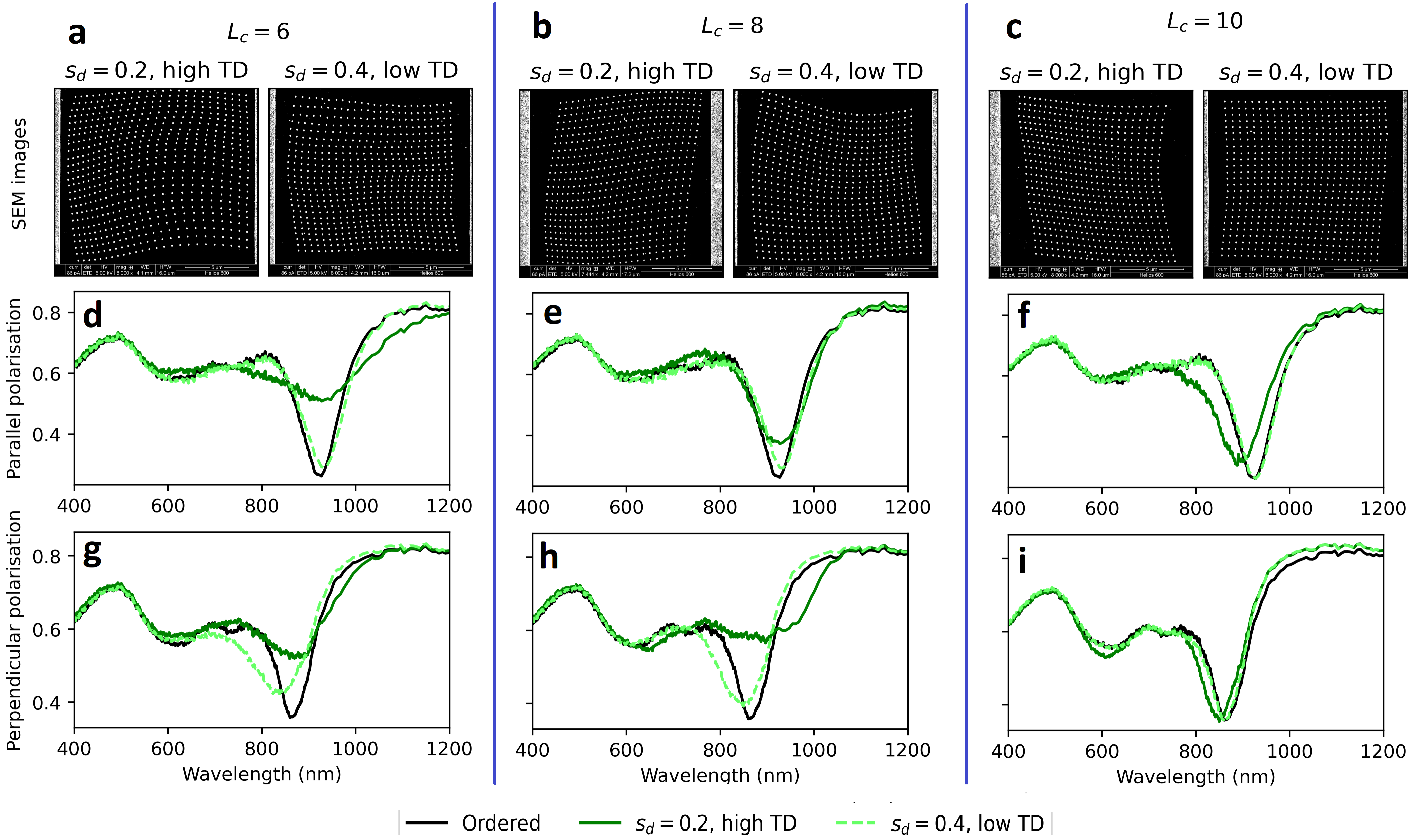} 
    \caption{SEM images of the experimental samples (top row) and their transmittance spectra under normal incidence light linearly polarised parallel (middle row) or perpendicular (bottom row) to the long axis of the nanostructures. Each plot displays the spectra of a low and high \td{} metasurface, dashed pink and solid red respectively, and an ordered metasurface with the same pitch (black).  Each column correspond to the metasurfaces generated with $\Lc\in [6,8,10]$ from left to right. 
    \label{fig:experiment}}
  \end{center}
\end{figure}

We additionally experimentally confirmed the link between \td{} and the strength of SLRs by designing metasurfaces  built using Focused Ion Beam (FIB) lithography.  Using three different correlation lengths $\Lc{} \in \{ 6,8,10 \}$ and starting from a regular square lattice of period \nm{500}, we generated several hundreds of lattices for two values of \Sd{}: 0.2 and 0.4.  For each value of \Lc{}, we selected two lattices to compare with each other: the one with the highest value of \td{} among those generated with \Sd{}=0.2 and the one with the lowest value of \td{} among those generated with \Sd{}=0.4.  Similarly to the previous section, we used \sh{} to select lattices of similar nearest neighbour distances.  We built two sets of seven metasurfaces, three pairs for each value of \Lc{} and one reference square lattice of period \nm{500}.  The two sets only differ in the size of the nanostructures, which in both cases were elongated \nm{50} thick gold nanodisk.  
The top nanodisk crossections are elliptical with $x$- and $y$-axis of size (160,180)~nm and (120,140)~nm for the first and second set respectively.
The resonant wavelength of the SLRs depends both on the distance between the nanostructures and on their polarisability.
The latter is strongly affected by the shape of the nanostructures and their anisotropy induces a shift of the SLRs wavelength of up to \nm{60} according to the polarisation of the exciting light.
We therefore report the optical properties of the metasurfaces excited under normal incidence light for two linear polarisation: polarised along the $y$-direction, parallel to the nanostructures' long axis and polarised along the $x$-direction, perpendicular to the nanostructures' short  axis.
SEM images of the  first set, as well as their transmittance spectrum compared to the square lattice are in figure~\ref{fig:experiment}. The results for the second set of metasurafaces, the comparison of these experimental results to the dipolar model and close up SEM images are in the supplementary information.

The three columns of figure~\ref{fig:experiment} contain for each \Lc, the SEM images of the designed pair of metasurfaces (first row) and their transmittance spectra upon excitation by light polarised parallel to the nanostructures' long axis (second row) and perpendicular to the nanostructures' short  axis (third row).  The transmittance spectrum of a periodic metasurface with the same pitch  is added for comparison (black lines).
We report in table~\ref{tab:Q_factor_vs_Sd_sv_Td} the quality factors of all the SLRs shown in figure~\ref{fig:experiment} as well as the \td{} of the corresponding metasurfaces.

\begin{table}[h]
    \centering
    \begin{tabular}{|c|c|c|c|}
    \hline
        Lattice parameters  & \td{}  & Q (parallel) & Q (perpendicular) \\ \hline
        \Lc{}=0, \Sd{}=0 & 0. & 10.1 & 11.5 \\ \hline
        \Lc{}=6, \Sd{}=0.2 & 0.030 & 4 & 5.2  \\ \hline
        \Lc{}=6, \Sd{}=0.4 & 0.012 & 9.3 & 6.8  \\ \hline
        \Lc{}=8, \Sd{}=0.2 & 0.025 & 6.7 & 4.4  \\ \hline
        \Lc{}=8, \Sd{}=0.4 & 0.005 & 7.8 & 7  \\ \hline
        \Lc{}=10, \Sd{}=0.2 & 0.026 & 8 & 11.5 \\ \hline
        \Lc{}=10, \Sd{}=0.4 & 0.002 & 10.1 & 11.5  \\ \hline
    \end{tabular}
    \caption{\td{} of the metasurfaces reported in  figure~\ref{fig:experiment} and the corresponding quality factors (Q) of their SLRs for parallel and perpendicular polarisation of the exciting light.}
    \label{tab:Q_factor_vs_Sd_sv_Td}
\end{table}

\begin{figure}
  \begin{center}
    \includegraphics[width=0.5\linewidth]{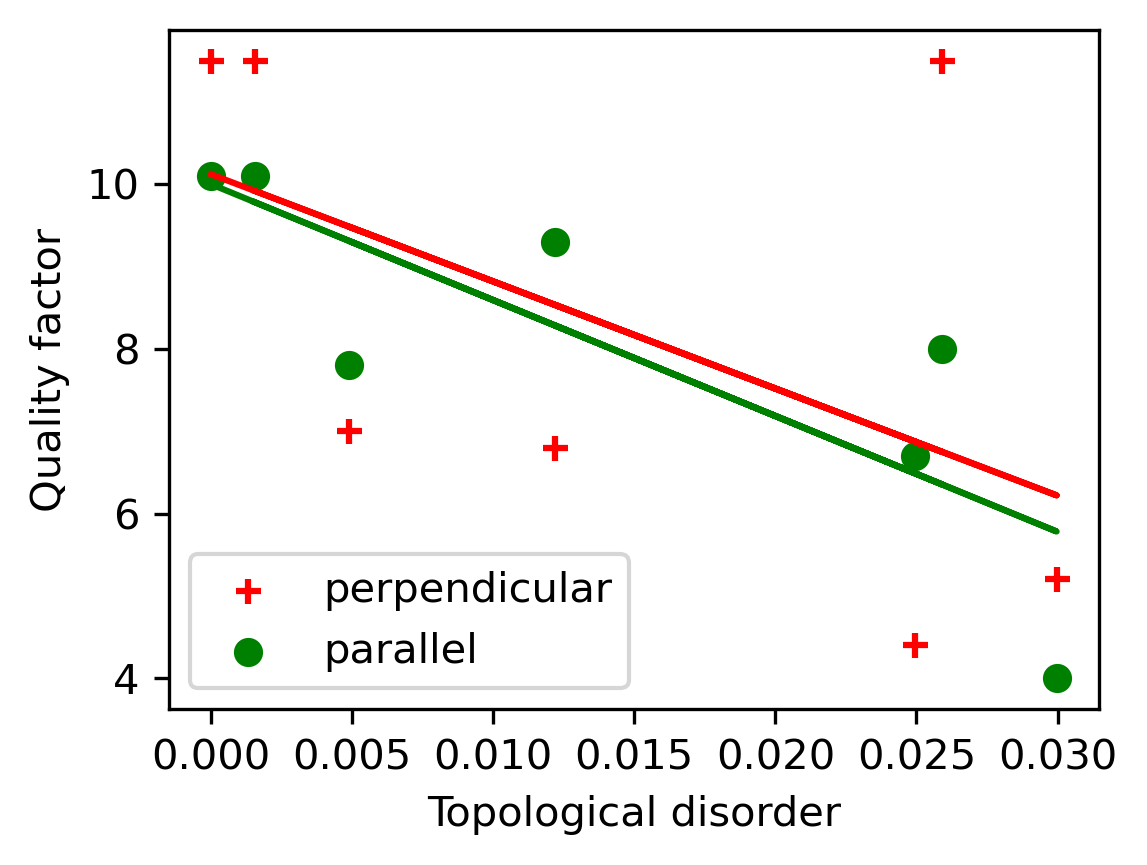} 
    \caption{Graph of the quality factors of the SLRs reported in figure~\ref{fig:experiment} in terms of the \td{} of the metasurfaces under normal incidence light linearly polarised parallel (green dots) and perpendicular (red crosses) to the long axis of the nanostructures.
    Two lines are added to represent the trend of the quality factors for the parallel polarisation, in green, and perpendicular polarisation, in red, in terms of \td{}.
    \label{fig:Q_vs_TD}}
  \end{center}
\end{figure}

As can be seen in figure~\ref{fig:experiment} and table~\ref{tab:Q_factor_vs_Sd_sv_Td}, in five configurations out of six, the SLRs of the metasurfaces designed with a high \Sd{} but a low \td{} are stronger and have a larger quality factor than the metasurfaces designed with a low \Sd{} but a high \td{}, proving that \td{} is indeed an accurate measure of disorder.
The only exception is the configuration with \Lc{}=10 and perpendicular polarisation, figure~\ref{fig:experiment}i, for which both metasurfaces have similarly strong SLRs with a quality factor of 11.5, equivalent to the square lattice for this polarisation.
Upon inspecting the lattices of the two metasurfaces generated for \Lc{}=10, shown in the panel c of figure~\ref{fig:experiment} or in larger versions in the supplementary information, one can visually appreciate that positional distortion seems to be only noticeable at a large-scale, while at short scales, the nanostructures seem to be more regularly spaced as if they were on a square lattice.
Indeed, larger values of \Lc{} average out the uncorrelated disorder of more nearby nanostructures which effectively smooths out the positional shift of each nanostructure, while maintaining large-scale shifts, responsible for the wavy patterns of the two right columns of figure~\ref{fig:Visual_ambiguity}.
Although \td{} is able to detect this large-scale distortion, and treats it similarly to short scale disorder, the strength of SLR depends mostly on the latter.
Indeed, the interaction strength between the nanostructures decreases as the inverse of the distance between them, under the dipolar approximation, and, effectively, a nanostructure only interact with a few of their neighbours. 
Therefore, a metasurface with short scale order but long scale disorder, such as those generated for \Lc{}=10 can exhibit strong SLRs despite having a large \td{}. 
This effect can also be visualised if we represent the quality factors of the SLRs in terms of the \td{} of the metasurfaces for both polarisation of the exciting light, figure~\ref{fig:Q_vs_TD}.
We see a decreasing trend of the quality factor in terms of \td{} despite outliers, such as the metasurface generated with \Lc{}=10, \Sd{}=0.2 that we just commented on, and some fluctuations that we similarly explain as \td{} being affected by large-scale disorder while the quality factor isn't.
A simple improvement would be to only define \td{} locally, on the scale that is relevant to the optical property dependant on the metasurfaces' disorder.
However, we chose for simplicity to keep the definition of \td{}, equation~\eqref{eq:TD}, global in this work.


This demonstrates that \td{} is a more accurate measure of the positional disorder of these metasurfaces compared to \Sd{} as in all of the cases reported here, the metasurface that should have been the most ordered, generated with the lowest value of \Sd{}, is actually at least as disordered as the metasurface that should have been the most disordered, generated with the highest value of \Sd{}.
Indeed, while a correlation length, induced by \Lc{}$\neq0$, made \Sd{} more ambiguous to describe the disorder of the metasurfaces, \td{} was able to accurately select lattices of chosen disorder, that we experimentally probed via the quality factor of their SLRs, despite the non unique relationship between \td{} and SLRs quality factors.

\section{Conclusion}

We have shown how Topological Data Analysis and persistent homology can be used to classify both correlated and uncorrelated disordered metasurfaces via their topological disorder.
In particular, we showed that for correlated disorder, topological disorder is a significantly more accurate measure of disorder than generative probabilistic parameters.
We proved, theoretically and experimentally, this accuracy by correlating topological disorder to the strength of surface lattice resonances of metasurfaces made of plasmonic nanostructures, despite the global definition of topological disorder being sensitive to large-scale distortion, while surface lattice resonances are not.
We argue that the universality, accuracy and computational speed of topological disorder makes it an advantageous tool to characterise and tune the fabrication methods of self assembled disordered metasurfaces, as well as to help design metasurfaces of specific degree of disorder, for example to enhance light extraction for more efficient LEDs or light absorption for improved solar cells.

\section{Experimental / Method}

The metasurfaces have lateral size approximately $12 \times \um{12}$ and were fabricated in a \nm{50} thick film of Au coated glass substrate using a focused ion beam facility, Helios Nanolab 600 from FEI ThermoFisher Scientific.  The metasurfaces were then spin-coated with IC1-200 whose refractive index is similar to that of the glass substrate. 

The spectral characterisation was performed in transmittance at normal incidence using a microspectrophotometer (CRAIC Technologies) equipped with a tungsten-halogen light source and cooled CCD array.  

The persistent homology of all lattices was computed using the Ripser python package\cite{Bauer2021Ripser}. The computation for each lattice, made of 625 nanostructures, was done in a fraction of a second.  The computation of the distance between each lattice's persistence diagrams considered for the figure~\ref{fig:Embedding} was done using the Wasserstein distance from the GUDHI python package\cite{gudhi:urm}.  Embeddings were obtained from the distance matrices by using classical multidimensional scaling.  We projected the embeddings in two dimensions for the visual representations in figure~\ref{fig:Embedding}.  
In general such embeddings live in a very high dimensional, non necessarily euclidean, space and a projection to a two dimensional flat space can lead to distortions. 
However, the magnitude of these distortions can be estimated in the classical multidimensional scaling methods by considering the relative absolute value of the eigenvalues of the embedding in each dimension\cite{cox_multidimensional_2000}. 
For the embedding represented in figure~\ref{fig:Embedding}, the eigenvalues of the two largest dimension, used to represent the embedding in 2D, are respectively 278 and 40 time larger than the largest negative eigenvalue, proving that an embedding in an euclidean space is a good approximation.
Similarly, the eigenvalues of the two largest dimension are respectively 22 and 3 times larger than the third largest positive eigenvalue, hinting that a projection in 2D is an accurate visual representation of the embedding. 

The numerical simulations of the metasurfaces optical properties were done using the discrete dipole approximation\cite{fradkin_fourier_2019} where each nanostructure is modelled as a dipole of the same polarisability. We assumed that the nanostructures were located in an homogeneous dielectric medium of refractive index $n=1.41$ which is a good approximation of the refractive index of the glass substrate and of the IC1 layer.  The reflectance was measured by computing the electromagnetic flux in the direction perpendicular to the surface, assuming a numerical aperture of $0.28$, to match the experimental setup.
The nanostructures polarisability was computed from simulating the optical response of an isolated nanostructure upon excitation by plane waves of different polarisability \cite{sterl_shaping_2021} , that we performed using the electromagnetic waves, frequency domain interface of the optics module of COMSOL 5.6, solved with a direct solver \cite{comsol}.

\begin{acknowledgement}

The authors acknowledge the use of the IRIDIS High Performance Computing Facility, and associated support services at the University of Southampton, in the completion of this work.
~
This work was supported by the Leverhulme Trust (grant RPG- 2019-055).

\end{acknowledgement}



\bibliography{Dis_meta}

\end{document}